\documentclass[useAMS,usenatbib,letters]{mn2e}

\usepackage{ads_refs}

\usepackage[english]{babel}
\usepackage{graphicx}
\usepackage{verbatim}
\usepackage{multirow}

\begin{document}
\title{Radio jets from stellar tidal disruptions}

\author[van Velzen, K\"ording and Falcke]{Sjoert van Velzen$^{1}$, Elmar K\"ording$^{1}$ and Heino Falcke$^{1,2,3}$ \\
$^{1}${Department of Astrophysics, IMAPP, Radboud University, P.O. Box 9010, 6500 GL Nijmegen, The Netherlands}\\
$^{2}$ASTRON, Dwingeloo, The Netherlands \\
$^3$ Max-Planck-Institut f\"ur Radioastronomie, Bonn, Germany }

\date{Received 2011 July 6; in original form 2011 April 20. Accepted 2011 July 12.  }
\pubyear{2011}

\maketitle 

\begin{abstract}
A star that passes too close to a massive black hole will be torn apart by tidal forces. The flare of photons emitted during the accretion of the stellar debris is predicted to be observable and candidates of such events have been observed at optical to \mbox{X-ray} frequencies.
If a fraction of the accreted material is fed into a jet, tidal flares should be detectable at radio frequencies too, thus comprising a new class of rare radio transients. 
Using the well-established scaling between accretion power and jet luminosity and basic synchrotron theory, we construct an empirically-rooted model to predict the jet luminosity for a time-dependent accretion rate. 
We apply this model to stellar tidal disruptions and predict the snapshot rate of these events. For a small angle between the observer and the jet, our model reproduces the observed radio flux of the tidal flare candidate GRB 110328A.
We find that future radio surveys will be able to test whether the majority of tidal disruptions are accompanied by a jet.
\end{abstract}

\section{Introduction}
When a star wanders too close to the massive black hole at the center of its galaxy, it will be tidally disrupted by the gravity of the hole \citep{Hills75}. 
After the disruption, about half of stellar mass remains bound \citep[e.g.,][]{Rees88, EvansKochanek89} and an electromagnetic flare is produced as the debris falls back onto the black hole. Theoretical efforts to predict this emission have focused predominately on the accretion of the bound stellar debris \citep[e.g.,][]{Rees88, LoebUlmer97, Ulmer99,Bogdanovic04} and the contribution from a super-Eddington outflow \citep{strubbe_quataert09,StrubbeQuataert10,LodatoRossi11}.

The electromagnetic flare from a stellar tidal disruption event (TDE) may be our only tool to probe dormant black holes, e.g., like the Galactic Center black hole \citep{MeliaFalcke01}, beyond the local universe  and could allow a much-anticipated study of black hole demographics as a function of galaxy type and cosmic time.
Individual TDE are also interesting since the sudden increase of accretion rate after the disruption, from zero to super-Eddington in a few months or even hours, is much more rapid than the timescale of normal accretion onto super-massive black holes. 

A number of (candidate) TDE have been identified in X-ray  \citep{Bade96, KomossaBade99, KomossaGreiner99, Esquej08} ---for a review see \citet{Komossa02}, UV \citep{Gezari06,Gezari08,Gezari09} and optical surveys \citep{vanVelzen10,Drake11, Cenko11}. Based on the optical luminosity of observed TDE one can predict that near-future synoptic surveys, such as LSST \citep{Ivezic08}, should detect thousands of such events per year \citep{Gezari09, vanVelzen10}. 

Follow-up observations of candidate TDE at radio frequencies is important to identify these events, as was realized when the first X-ray candidates were detected \citep{Komossa02}. However, blind radio variability surveys also have the potential to \emph{discover} TDE. The rapid increase of sky coverage and sensitivity of variability surveys at both high and low frequencies promises an exciting future for this field.

Recently, \citet{GianniosMetzger11} proposed a model for the radio emission from TDE, based on the interaction of the jet with the interstellar medium (ISM); \citet{Bower11} compared their predictions to upper limits of existing radio surveys for transients. 
The present work is an extension of the approach outlined in \citet*{vanVelzen10b}, where we use the well-established jet-disk symbiosis model to calculate a time-dependent jet model for TDEs. 
We will only consider the emission from the compact core of the jet; interactions with the surrounding medium may enhance the jet luminosity in some cases, but here we aim to obtain a conservative model and we therefore consider solely the internal jet emission. 

In section \ref{sec:jetlum} we present our time-dependent jet model. In section \ref{sec:lc} we discus the light curves produced by our model and compare them to radio observations of candidate TDE  such as GRB 110328A. 
We predict the snapshot rate of jets from TDE in section \ref{sec:snap} and compare this rate to the sensitivity of current and future radio transient surveys.

\section{Time-dependent jet model}\label{sec:jetlum}
There is already quite a range of time-dependent models for radio jets in super-massive black holes in the literature \citep[e.g.,][]{ChiabergeGhisellini99,Gupta06}, but they typically only address relatively small changes or focus on a subclass of AGN. A major question remains how jets evolve, as a function of time when accretion suddenly sets in and increases by many orders of magnitude. There is increasing consensus that accretion discs and jets are intrinsically coupled and are best understood as a symbiotic system. Evidence that jets are ubiquitous to accretion comes from the `fundamental plane of black holes', which provides a universal scaling law for the non-thermal emission of black holes over all mass scales \citep*{Merloni03,Falcke04}. 
We thus proceed under the hypothesis that all accreting massive objects, including TDEs, launch a jet, but, as discussed later, take potential radio-loud/radio-quiet switches at high accretion rates into account. 

 In this section, we will first generalize the jet-disk symbiosis model of \citet[][FB95, hereafter]{falcke95I} to a time-dependent accretion rate and we then apply this model to TDE. 

\subsection{Basic jet model}
The essence of jet-disk symbiosis is power unification: $Q_j = q_j L_d \propto q_j\dot{M}$, the jet power ($Q_j$) is some fraction ($q_j$) of the disk luminosity ($L_d$), which is a linear function of the accretion rate ($\dot{M}$). If we assume equipartition between the energy in relativistic particles and the magnetic field, the synchrotron emissivity follows from the accretion rate: $\epsilon_{\rm syn} \propto B^{3.5} \propto (q_j\dot{M})^{1.75} z^{-3.5}$, with $z$ the distance to the origin of the jet (FB95, Eq. 19). We obtain the synchrotron luminosity of the jet ($L_\nu$) by integrating the emissivity over the jet volume, a cylindrical-symmetric cone,
\begin{eqnarray}\label{eq:jd}
  L_\nu  = C_{\rm eq} \int_{z_{\rm ssa}}^{\infty}\,{\rm d}z\,  z^2 \epsilon_{\rm syn}(z,\nu/\delta) \propto (q_j L_d)^{17/12}
\end{eqnarray}
(FB95, Eqs. 52 \& 56). Here $\delta$ is the Doppler factor of the jet and $\nu$ is the observed frequency. The lower limit of integration, $z_{\rm ssa}(\nu/\delta)$, is the distance where the jet becomes optically thin to synchrotron self-absorption. The normalization ($C_{\rm eq}$) is the conversion factor between jet power and jet luminosity, which can be estimated using equipartition arguments or obtained by observations. 

The great success of jet-disk symbiosis is that the observed properties of all accreting black holes with radio-loud jets can be fit with $q_j=0.2 \equiv q_{\rm loud}$ and a single value of the normalization ($C_{\rm eq}$) of Eq. \ref{eq:jd} \citep{falcke95II, Kording08}. In his work, we will fix $C_{\rm eq}$ using the empirical normalization found by \citet{Kording08} for efficient accretion, $L_d = 0.1 c^2 \dot{M}$, and we will use a jet Lorentz factor $\gamma_j=5$ \citep{falcke95II} throughout. 

The ``classic'' jet model (Eq. \ref{eq:jd}) is derived for a constant accretion rate; to use this model for a time-dependent accretion rate, $\dot{M}(t)$, we have to consider three things: (i) the non-zero time delay of photons emitted at different locations in jet, (ii) $z_{\rm ssa}(t)$ will set the time scale of the emission, and (iii) the emissivity becomes a function of time. The latter of these changes is trivial to apply because at the base of the jet, the relation between the synchrotron emissivity and accretion rate is given by the standard jet-disk model and all one has to do is to propagate $\epsilon_{\rm syn}$ forward in time using $z(t) = t \beta_j c$. To account for (ii) we use $\tau \propto z \kappa_{\rm syn} / \sin(i)=1$, where $\kappa_{\rm syn}$ is the synchrotron emission coefficient, to find
\begin{equation}\label{eq:z_ssa}
z_{\rm ssa} = 1\, {\rm pc}\; f\frac{\mbox{GHz}}{\nu/\delta} \left(\frac{q_j(t)}{0.2}\frac{L_d(t)}{10^{45}\, {\rm erg}\,{\rm s}^{-1}}\right)^{\frac{2}{3}} \left(\frac{\beta_j}{\sin(\frac{i}{30^{\circ}})} \frac{5}{\gamma_j}\right)^{\frac{1}{3}}
\end{equation} 
(FB95, Eq. 52), here $f\sim1$, is a factor that depends on the details of equipartition. We preform a check on the latter using observations of NGC 4258 at 22 GHz showing the base of the jet at a minimum distance of $0.012$ pc from the dynamical center of the accretion disk \citep{Herrnstein97}; using $i_{\rm obs}=83^{\circ}$ and $\gamma_j=3$ \citep{Yuan02} at the base of the jet and $\dot{M}=0.01 M_\odot\,{\rm yr}^{-1}$ \citep{Gammie99}, we obtain $f\approx 0.5$. 

We can now modify the integral of Eq. \ref{eq:jd} to obtain our time-dependent jet model, 
\begin{equation}\label{eq:jd_time}
L_\nu(t) = C_{\rm eq} \delta^2 \int_0^{z_{\rm dec}} {\rm d}z\, z^2 \epsilon_{\rm syn}(t_r, z,\nu/\delta) \Theta_{\rm ssa}(t_r,z,\nu/\delta)\quad.
\end{equation}
Here $\Theta_{\rm ssa}(t,z,\nu)$ is a step function that enforces a crude radiative transfer: it is zero for $z<z_{\rm ssa}(t)$ and unity for $z>z_{\rm ssa}(t)$. The retarded time, $t_r$, is introduced to ensure that we integrate using only the photons that will arrive simultaneously at the observer, $t_r(t,z) = t - z \cos(i)\, c^{-1} $with $i$ the angle between the jet and observer, in the rest-frame of the jet. Note that for observed angles $\cos(i_{\rm obs})<\beta_j$, we have $t_r>t$; the photons from the middle of the jet arrive simultaneous with photons emitted further ahead, i.e., the jet appears to be seen from behind in the observer frame \citep[e.g.,][]{Jester08}.
The upper limit of integration, $z_{\rm dec}$, is the radius where the jet will slow down significantly because the initial jet energy equals the energy of the shocked matter swept up  by the jet \citep[e.g.,][]{Piran04}: $z_{\rm dec}\propto (E_j/n\gamma_j^2)^{1/3}$, with $E_j \propto  \int q_j \dot{M}\,{\rm d}t$ and $n$ the ISM number density. 
In the following section we discus $q_j(t)$ and $\dot{M}(t)$ for tidal disruption events and give the physical scale of $z_{\rm dec}$.


\subsection{Accretion states of TDE}\label{sec:acc}
To apply the time-dependent jet-disk symbiosis model (Eq. \ref{eq:jd_time}) to TDE we need the accretion rate as a function of time and black hole mass. We first consider the time it takes for most of the stellar debris to return to the pericenter ($R_p$) after the disruption,  $t_{\rm fallback} \sim 0.1 (M_{\rm BH} / 10^6 M_\odot)^{1/2} (R_p/R_t)^{3}\, {\rm yr}$ for a solar-type star \citep[e.g.,][Eq. 3]{Ulmer99}, $R_t$ is the tidal disruption radius. After this time, the material falls back onto the black hole at a rate, 
$\dot{M}_{\rm fallback} \approx 1/{3}\, {M_*}/{t_{\rm fallback}}({t}/{t_{\rm fallback}})^{-5/3} $
 \citep{Rees88}, here $M_*$ is the mass of the star. We will use $\dot{M}_{\rm fallback}$ with $R_p=R_t$ for the accretion rate onto the black hole that can be fed into the jet. For $M_{\rm BH} < {\rm few} \, 10^{7} M_\odot$, the fallback rate will (greatly) exceed the Eddington rate for some time, but we will conservatively asume that $\dot{M}(t)=\dot{M}_{\rm Edd}$ during this time; we use an exponential rise to the peak accretion rate for $t<t_{\rm fallback}$. Our results are not sensitive to potential deviations from the canonical $t^{-5/3}$ scaling of the fallback rate \citep*[e.g.,][]{Lodato09}, because most of the energy is injected into the jet during the super-Eddington phase, where $\dot{M}$ is capped at $\dot{M}_{\rm Edd}$. 

With the accretion rate given by the theory of tidal disruptions, we only have to provide one more ingredient to produce radio light curves for these events: the fraction of accretion power that is fed into the jet. 
Jets from active super-massive black holes can be radio-loud or radio-quiet \citep{Kellermann89}, which appears to be at odds with jet-disk symbiosis. However, detailed observations have shown that nearly all radio-quiet AGN do show some radio emission which can be interpreted as originating from the core of a relativistic jet \citep{Brunthaler00,Falcke01}. Indeed radio-quiet jets can also be accommodated by Eq. \ref{eq:jd} by reducing $C_{\rm eq}$ or $q_j$ with a factor $\sim 10^2$ with respect to radio-loud systems.

If we assume that the physics behind launching the jet and producing the synchrotron emission is no different for TDE and ``normal'' active black holes, we are left to answer the following question: \emph{is a TDE jet radio-loud or radio-quiet?} 
Observations of accreting stellar mass black holes (i.e., X-ray binaries) can help to answer this question since they are variable on timescales down to minutes \citep[][]{Belloni05} and they can serve as examples for AGN \citep{McHardy06,Chatterjee11}. 

When X-ray binaries experience a burst of accretion, they follow a  predefined track in the hardness-intensity diagram \citep{Belloni05} corresponding to distinct accretion states with associated jet properties \citep{Fender04}. In the quiescent mode (the hard-state) and during the onset of the burst, jets in X-ray binaries are radio-loud, while in the high-accretion mode (the soft-state) they are radio-quiet. 

The sudden enhancement of the accretion rate during a TDE, may move it through the different modes of accretion in two ways: directly into the radio-quiet soft-state, or into the soft-state via the radio-loud burst-state. Alternatively, the jet from a TDE may behave like a radio-loud quasar at all times. We therefore consider three different scenarios for the fraction of accretion energy that is fed into the jet:
\begin{eqnarray}\label{eq:scen}
  q_j = \left\{
  \begin{array}{l l c}
    q_{\rm loud}& \quad {\rm all~times} & (a) \\
    q_{\rm loud}/10^2& \quad \dot{M}(t)>2\% \dot{M}_{\rm Edd} &(b)\\
    q_{\rm loud}&\quad t<t_{\rm fallback} & (c) \\
      \end{array} 
    \right..
\end{eqnarray}
where each scenario reverts to the preceding one if the condition on $t$ or $\dot{M}$ is not true (e.g., $q_j=q_{\rm loud}=0.2$ if $\dot{M}<2\%\dot{M}_{\rm Edd}$ in all three scenarios).
In scenario $b$  the jet becomes radio-loud only when the accretion drops below $<2\% \dot{M}_{\rm Edd}$ \citep{Maccarone03}, while in scenario $c$ the systems makes a full loop trough all accretion modes, starting with a radio-loud burst during the onset of the accretion.  We consider $a$ most optimistic, $b$ most pessimistic, and $c$ the most likely scenario. The two orders of magnitude difference in $q_j$ between scenarios $a$ and $b$ can also be taken to reflect our uncertainty on the coupling between jet power and accretion during the super-Eddington phase of the disruption. 

With $q_j$ and $\dot{M}$ at hand, we can now calculate, $z_{\rm dec}$, the radius where the jet will slow down significantly, which is the upper limit of the integral over jet volume (Eq. \ref{eq:jd_time}). For $M_{\rm BH}=10^{7}\,M_\odot$ and scenario $a$ (Eq. \ref{eq:scen}), using a jet opening angle of $7^{\circ}$ (FB95) and an ISM density of 1 proton per ${\rm cm}^{-3}$, we obtain $z_{\rm dec} =3.5\, {\rm pc}$. Comparing this to $z_{\rm ssa}$ (Eq. \ref{eq:z_ssa}) this implies a significant suppression of the luminosity for observers looking at $\nu<500\, {\rm MHz}$ because $z_{\rm dec} < z_{\rm ssa}(\nu)$. However, this suppression is less relevant at  lower $M_{\rm BH}$ or $q_j$, since $z_{\rm dec}\propto (q_j L_d)^{1/3}$ while $z_{\rm ssa}\propto (q_j L_d)^{2/3}$. Clearly, the density distribution within a few parsec from the black hole varies between galaxies: each TDE jet will face a different deceleration radius. In elliptical galaxies, $z_{\rm dec}$ is likely to be larger by at least a factor $10$ with respect to the value adopted in this work, due to the low gas density in these galaxies \citep[e.g.,][]{Biermann83}. On the other hand, $z_{\rm dec}$ can decrease if the jet runs into a high-density clump of matter, which will enhance the luminosity, as seen in an exemplary way in the radio-intermediate quasar \mbox{III Zw 2} \citep{Brunthaler00}. For galaxies where $z_{\rm dec}<0.1\, {\rm pc}$, the external emission as modelled by \citet{GianniosMetzger11} dominates over emission from the core of the jet at all relevant frequencies. Discriminating between core and external emission for individual TDE jets may be possible using the time delay between the radio emission and the time of disruption.


\begin{figure}
  \begin{center}
    \includegraphics[trim=6mm 2mm 6mm 4mm,clip, width=.45 \textwidth]{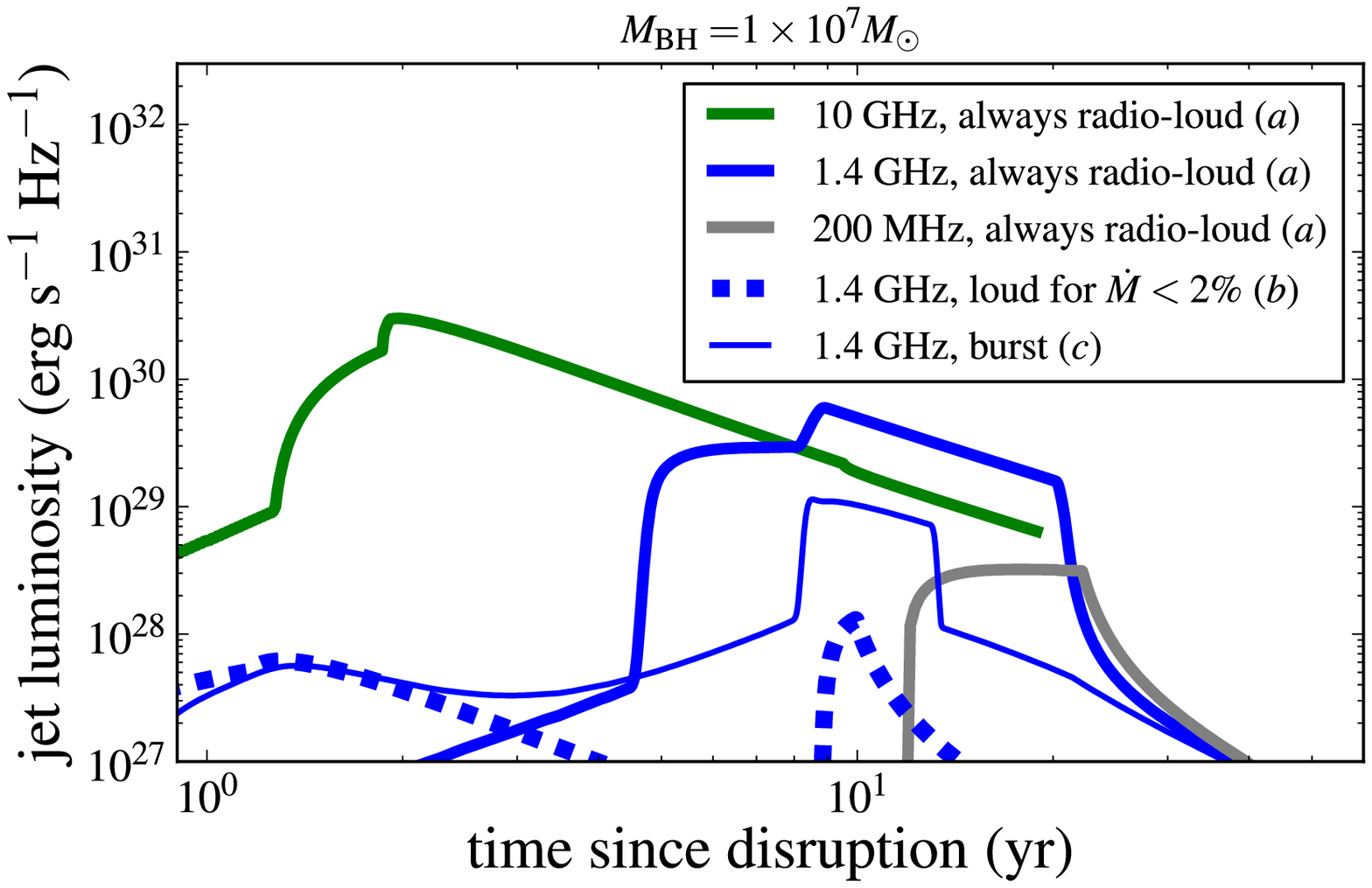} 
    \caption{Light curves for synchrotron emission for jets from TDE for $i_{\rm obs} = 30^{\circ}$, $M_{\rm BH}= 10^7\,M_\odot$ and three different scenarios of coupling between accretion and jet power ($a$, $b$ and $c$ in the legend refer to Eq. \ref{eq:scen}). For the ``always radio-loud'' scenario, we show three different frequencies (thick solid lines). 
The highest frequencies are visible at the earliest times and at highest luminosity because $z_{\rm ssa}\propto \nu^{-1}$ (Eq. \ref{eq:z_ssa}). 
For the ``burst'' scenario (thin line) we see a strong luminosity increase corresponding to the radio-loud part of the jet during the start of the accretion, as expected, this peak coincides with the peak of scenario $a$.}\label{fig:lc_bh}
  \end{center}
\end{figure}

\section{Radio light curves}\label{sec:lc}

In Fig. \ref{fig:lc_bh} we show the radio light curves that result from applying the jet-disk symbiosis to TDEs. For the scenario in which the jet is always radio-loud (Eq. \ref{eq:scen}$a$), one can see most clearly how the opacity sets the timescale of the emission. Since $z_{\rm ssa} \propto \nu^{-1}$ (Eq. \ref{eq:z_ssa}), the jet is visible at earlier times and at higher luminosity for higher frequencies. 
The sudden drop in luminosity after about $20$~yr is caused by our fixed upper limit of Eq. \ref{eq:jd_time} ($z_{\rm dec}$): we stop following the jet beyond this point because the aim of this work is to predict the internal jet emission. Connecting the internal and external emission in a single model will be the subject of future work. 
At $\nu=200$ MHz, we see a plateau of constant luminosity which is caused by the photons produced after the super-Eddington phase. 
For a given black hole mass, the duration of the radio flare is maximal if viewed along the critical angle, $i_{\rm obs} = \arccos(\beta_j)$; within this angle, the timescale is shorter because most photons arrive nearly simultaneously at the detector, while at larger viewing angles, the frequency in the rest-frame of the jet ($\nu/\delta$) increases, making the jet visible at earlier times. 

\begin{figure}
  \begin{center}
    \includegraphics[trim=10mm 2mm 8mm 9mm,  width=.45\textwidth]{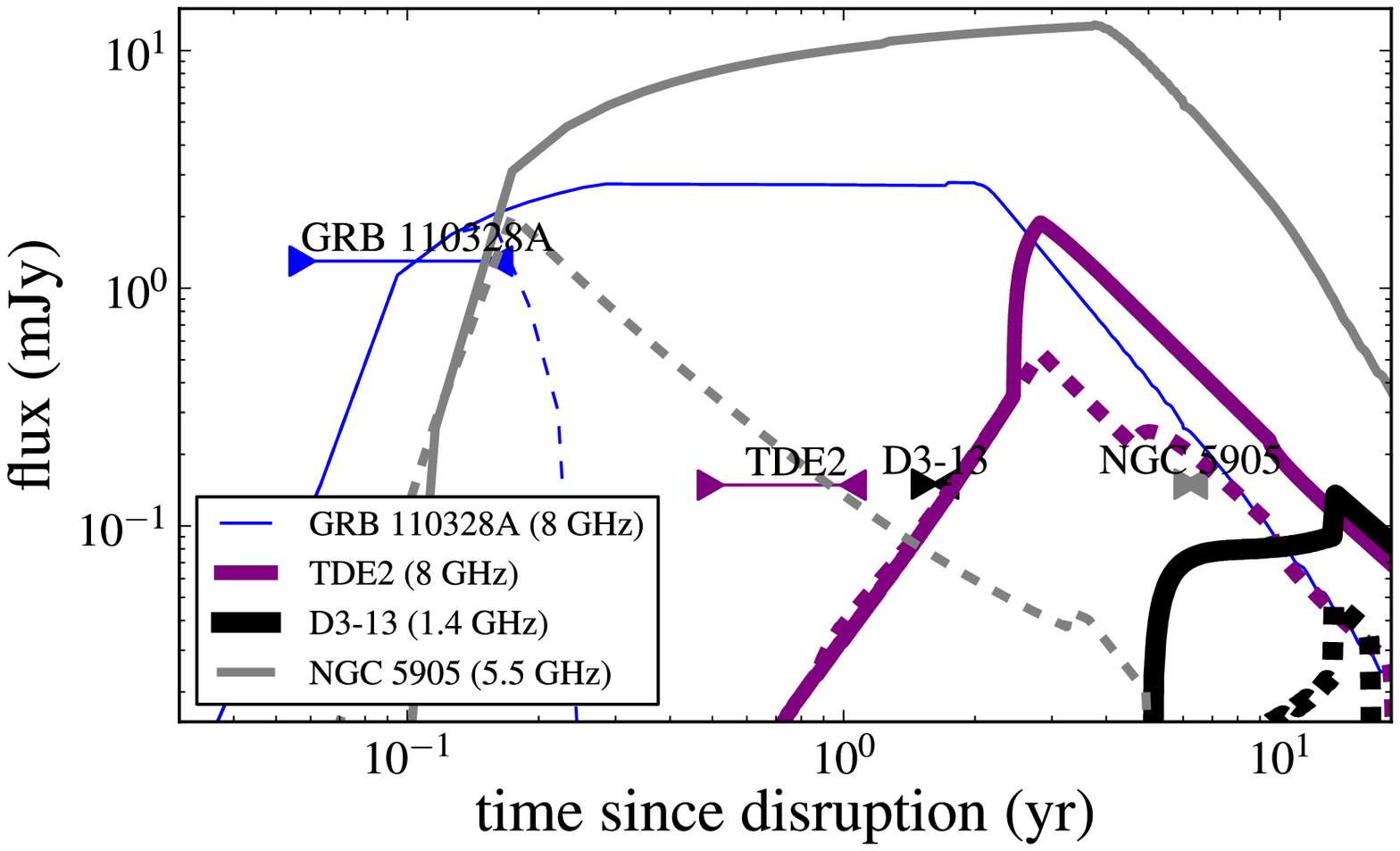} 
    \caption{The predicted flux for TDE2 $M_{\rm BH} \sim 5\times 10^{7}\,M_\odot$ \citep{vanVelzen10}, D3-13, $M_{\rm BH} \sim 2\times 10^{7}\,M_\odot$ \citep{Gezari08}, the X-ray flare from NGC 5905, $M_{\rm BH}\sim 2\times 10^5\, M_\odot$ \citep{KomossaBade99} and GRB 110328,  assuming $M_{\rm BH} \sim 1\times 10^{6}\,M_\odot$. We show our most optimistic model $a$ (solid line) and the more realistic ``burst model'' $c$ (dashed line). We use $i_{\rm obs}=30^{\circ}$ for the first three candidates and show the \mbox{3-$\sigma$} upper limits at $\nu=8$ GHz \citep{vanVelzen10}, $\nu=1.4$ GHz \citep{Bower11} and $\nu=8$ GHz \citep{Komossa02} from the last radio observations. For GRB 110328, we use $i_{\rm obs}=1^{\circ}$ and we show the VLBA detection at 8.4 GHz \citep{Levan11}. The triangles pointing right and left correspond to the lower and upper limit on the time of disruption, respectively. }\label{fig:radio_obs}
\end{center}
\end{figure}

In Fig. \ref{fig:radio_obs} we show follow-up radio observations that have been obtained for some candidate TDE. The upper limits on the radio luminosity are consistent with our most optimistic prediction for the jet luminosity, except for the candidate in NGC 5905 which is only consistent with scenarios $b$ and $c$. We note that observations of similar depth obtained today, $\sim 5$ years after the flare, should yield a detection. Finally we consider the recently discovered GRB 110328A / Swift J164449.3+573451, which may be an example of a strongly beamed TDE \citep[e.g.,][]{Bloom11, Levan11,Zauderer11}; indeed for $i_{\rm obs}<10^{\circ}$ and $M_{\rm BH} = 10^6 M_\odot$ our model with scenarios $a$ yields the observed VLBA radio flux \citep{Levan11} of this transient. If we conservatively assume that the first Swift detection marks the start of the disruption, our model requires  $i_{\rm obs}<1^{\circ}$ to explain the $\sim {\rm days}$ delay between gamma-ray and radio photons; this angle constraint becomes less stringent if the high-energy photons originate from the jet.

\section{Snapshot rate}\label{sec:snap}
Using the model presented in section \ref{sec:jetlum}, we can predict how many jets are visible above a certain flux limit ($F_{\rm lim}$) at any moment in time,
\begin{eqnarray}
  N(F_{\rm lim}, \nu) &=& (4\pi)^{-1} \dot{N}_{\rm tde} \, \int {\rm d}\Omega_{\rm obs} \int{\rm d}z\, 4\pi d_C^2(z) \times \nonumber  \\
 && \int {\rm d}M_{\rm BH}\, \,\phi_{\rm BH}\, \tau_{\rm eff}(L_\nu,d_L(z),F_{\rm lim}) \; . \label{eq:snap}
\end{eqnarray}
Here $d_C(z)$ and $d_L(z)$ are the co-moving and luminosity distance\footnote{We adopt a standard cosmology with $H_0 = 72\, {\rm km}\,{\rm s}^{-1}{\rm Mpc}^{-1}$, $\Omega_m=0.3$ and $\Omega_\Lambda=0.7$.}, respectively and $\phi_{\rm BH}$ is the black hole mass function. The integration over viewing angles, ${\rm d}\Omega_{\rm obs}$, accounts for the effects of Doppler boosting. Finally, our jet model enters via $\tau_{\rm eff}(L_\nu(M_{\rm BH}, i_{\rm obs}), d_L, F_{\rm lim})$ or the ``effective time'' given by the part of the light curve that obeys $L_\nu(t) / (4\pi d_L^2)>F_{\rm lim}$. 
We also consider the model by \citet{GianniosMetzger11} using their Eq. 8, with fiducial parameters.

We use the local black hole mass function of \citet{Marconi04} for $\phi_{\rm BH}$ and a TDE rate per black hole of $\dot{N}_{\rm tde}=10^{-5}\,{\rm yr}^{-1}$ which is based on the observed rate per galaxy from SDSS observations \citep[$3\times10^{-5}\,{\rm yr}^{-1}$,][]{vanVelzen10} and ROSAT observations \citep[$9\times 10^{-6}\,{\rm yr}^{-1}$,][]{donley02}. At the lowest flux limit we consider, $F_{\rm lim}=0.05\, {\rm mJy}$, $\tau_{\rm eff}(z)\times d_c^2$ peaks at $z=0.5$ so we are not sensitive to cosmological evolution of $\phi_{\rm BH}$ or $\dot{N}_{\rm TDE}$. Since $L_\nu$ peaks at $M_{\rm BH}\sim 5\times 10^{7}\,M_\odot$ and $\phi(M_{\rm BH})$ flattens towards low black hole mass, Eq. \ref{eq:snap} is not sensitive to the upper or lower boundaries of the integration over black hole mass.

\begin{figure}
\begin{center}
\includegraphics[trim= 10mm 2mm 8mm 9mm,  width=.45\textwidth]{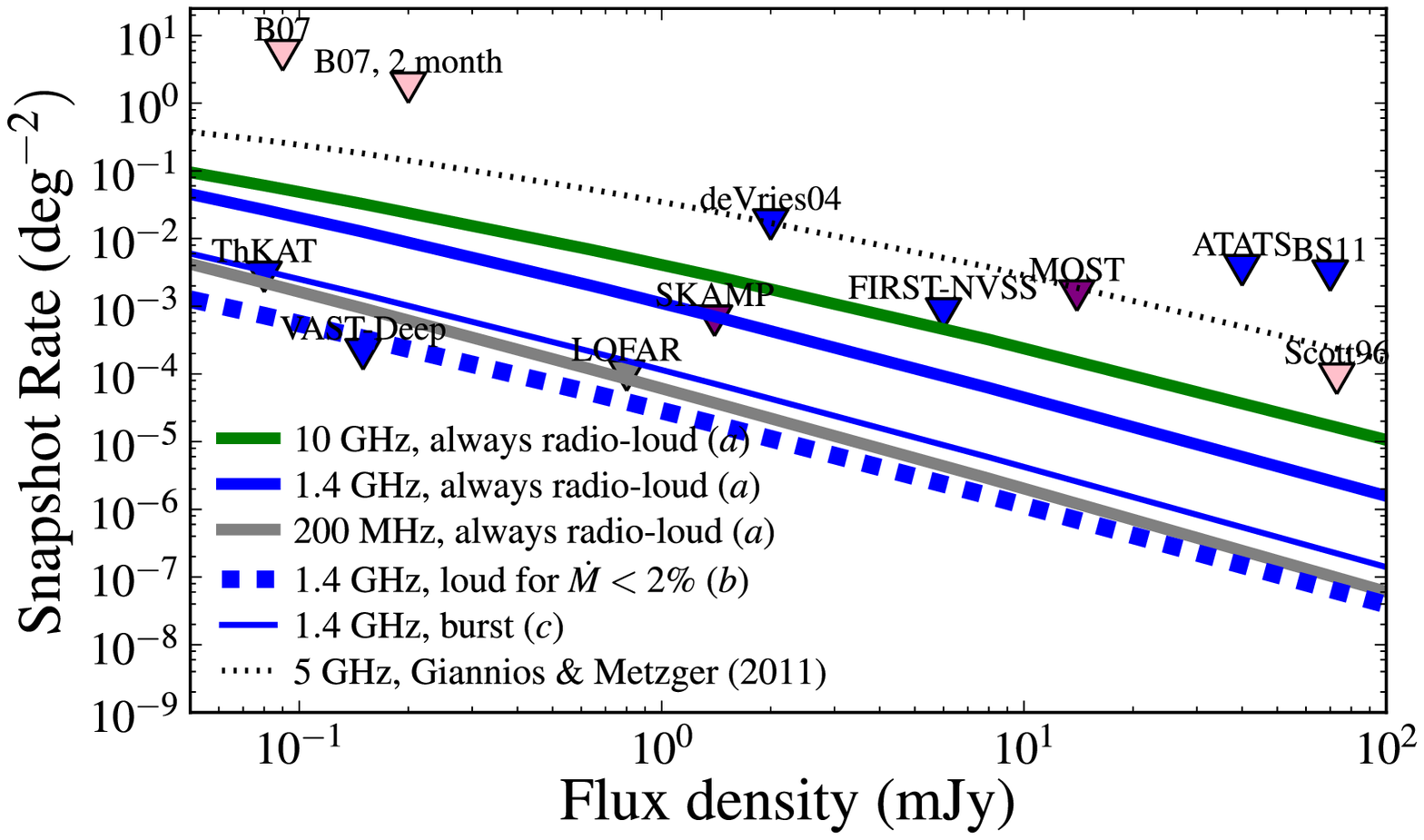} 
\caption{The snapshot rate of TDE jets. We show \mbox{2-$\sigma$} upper-limits from: \citet{Scott96} and \citet[][B07]{Bower07} at 5 GHz, \citet[][FIRST-NVSS]{Levinson02, Gal-Yam06}, \citet{deVries04}, ATAS \citep{Croft10} and \citet[][BS11]{BowerSaul11} at 1.4 GHz, MOST \citep[][]{Bannister11} at 843 MHz. We also show the limits that can be obtained if no candidates are detected in (near) future variability surveys. We refer to \citet{Ofek11} for an overview of radio variability surveys.  }\label{fig:snap}
\end{center}
\end{figure}

In Fig. \ref{fig:snap} we show the snapshot rate for the three different scenarios we consider (Eq. \ref{eq:scen}) and three different frequencies. At higher frequencies the jets are brighter and thus visible out to a larger volume, while at lower frequencies the duration is longer.  
These competing effects also imply that any uncertainty on $z_{\rm ssa}$ (Eq. \ref{eq:z_ssa}) has limited influence on the predicted snapshot rate. We also compare our predicted snapshot rate to observed upper limits on the rate of extra-galactic radio transients. For surveys with detected transients, we use the classification by \citet{Bower11} to limit these radio transients to potential TDE jets only.

The current radio transient surveys are not sensitive or large enough to test our prediction of the snapshot rate. This changes, however, when we consider the potential of near-future projects. 
For LOFAR\footnote{www.lofar.org}, we use 0.25~mJy for the thermal rms obtained at 180 MHz in a survey that will cover $2\pi$ steradian in about 3 months 
We also consider SKAMP\footnote{www.physics.usyd.edu.au/sifa/Main/SKAMP},
ThunderKAT, which is part of MeerKAT\footnote{www.ska.ac.za/meerkat}
and the VAST project, which is part of ASKAP\footnote{www.csiro.au/science/ASKAP}.
Using 3 times the rms for the detection threshold, we find that for the optimistic scenario (Eq. \ref{eq:scen}$a$), the SKAMP, LOFAR, surveys should contain about 2 jets from TDE. The VAST project is sensitive enough to test even the most conservative scenario $b$.

\section{Summary \& Conclusion }
We have presented a time-dependent jet-disk symbiosis model that yields a conservative and robust estimate of the radio luminosity of the compact jet that likely accompanies stellar tidal disruption events. This model is consistent with current constraints of the radio properties of TDE candidates and naturally predicts the observed radio flux of the newly discovered GRB 110328A. Based on our predicted snapshot rate we conclude that future radio surveys will be able to test whether the majority of tidal disruptions are indeed accompanied by a relativistic jet.

{\small
\subsection*{Acknowledgements}
We thank the anonymous referee for comments that improved the manuscript, A. Marconi for sharing data of $\phi_{\rm BH}$ and D. Frail, D. Giannios,  S. Komossa, and R. Wijnands for useful discussions.
}


\end{document}